\documentclass[amsmath, amssymb, aps, prb, superscriptadress]{revtex4-1}

\usepackage{graphicx}
\usepackage[colorlinks]{hyperref}
\usepackage[caption=false]{subfig}
\usepackage{bm}
\usepackage{wrapfig}

\begin{document}

\title{Fermi liquid theory applied to a film on an oscillating substrate}
\author{J. A. Kuorelahti, J. A. Tuorila and E. V. Thuneberg} 
\affiliation{Department of Physics,
P.O.Box 3000, FI-90014 University of Oulu, Finland}

\date{\today}

\begin{abstract}
We consider a film of a normal-state Fermi liquid on a planar substrate. 
Landau's Fermi liquid theory is applied to calculate the linear response of the film to transverse oscillation of the substrate. The response consists of a collective transverse zero sound mode, as well as incoherent quasiparticle excitations of the degenerate fermions. We calculate numerically the acoustic impedance of the film under a wide range of conditions relevant to normal state \textsuperscript{3}He at millikelvin temperatures. Some cases of known experiments are studied but most of the parameter range has not yet been tested experimentally.

\end{abstract}

\maketitle

\section{Introduction} \label{intro}

\begin{figure*}[bt]
        \centering
                \includegraphics[width=0.8\textwidth]{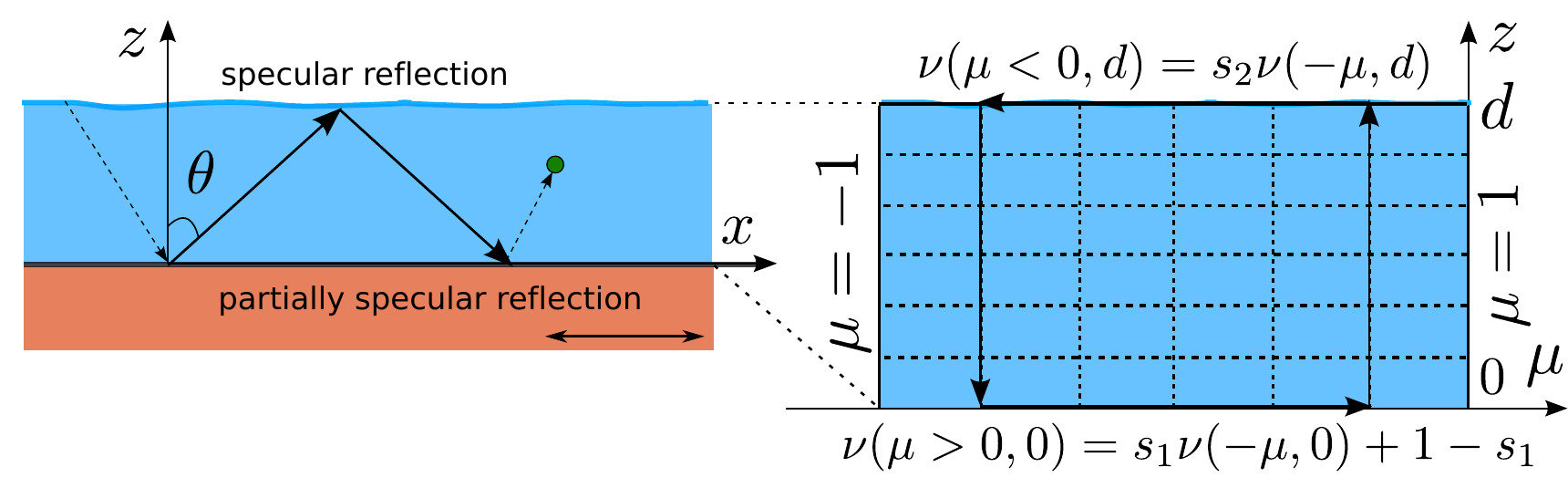}
        \caption{Simplified diagram of the problem of a liquid film on an oscillating substrate, and its solution in the space formed by momentum projection $\mu=p_z/p_F=\cos\theta$ and the dimensionless vertical coordinate $\zeta=z/d$. The boundary conditions are given in Eq.\ \eqref{e.bcon}.}\label{fig:FLtrajec}
\end{figure*}

Let us consider a layer of liquid on a planar substrate. Assume the substrate oscillates harmonically in its plane. The liquid is dragged into motion by the moving substrate. A measurable quantity is the transverse acoustic impedance of the liquid layer. It is defined as the ratio of the force on the liquid to the velocity amplitude of the substrate. The impedance consists of a dissipative real part and a reactive imaginary part. The latter can be interpreted as the amount of mass of the liquid that is coupled to the oscillation of the  substrate. For an ordinary liquid the Navier-Stokes equations reduce to a diffusion equation and the transverse acoustic impedance can straightforwardly be calculated. A schemantic of the set up is presented in Fig.\ \ref{fig:FLtrajec}.

The purpose of this paper is to calculate the transverse acoustic impedance
of a layer of a Fermi liquid. By Fermi liquid we mean that the fluid is described by Landau's Fermi liquid theory \cite{Landau57}. Landau's theory is a paradigm of what can be the state of an interacting many-body system.  The central idea is that although the particles are strongly interacting, the low-energy properties of the system can be described by weakly interacting excitations called quasiparticles. Similar to molecules in a rarefied gas, the quasiparticles can have a long mean free path, but there is an essential difference that even in the absence of collisions, interactions between the quasiparticles remain. This has important effects. For example, it allows the propagation of density oscillations even in the absence of collisions, so called zero sound. Also transverse oscillations can propagate as a wave, in contrast to Navier-Stokes fluid where such motion obeys a diffusion equation. Landau's theory is explained in many articles and textbooks \cite{LLs2,Abrikosov59,Baym91,Leggett75}. Originally Landau formulated the theory for liquid \textsuperscript{3}He, but it also forms the basis for understanding the behavior of conduction electrons in metals. Extension of the Fermi liquid theory to include paring correlations gives an accurate description of the superfluid or superconducting state of a Fermi liquid \cite{Leggett75,Serene83}. 

The calculation of the impedance requires solution of the transport equation, the Landau-Boltzmann equation in appropriate geometry. This was first done by Bekarevich and Khalatnikov \cite{Bekarevich61}. Their solution was extended by
Flowers and Richardson \cite{Flowers78}. These solutions are basically analytic but they are very complicated. Simpler approximate results were derived by Richardson\cite{Richardson78}. All these assume a thick liquid layer, in principle filling a half space. Our purpose is to generalize these calculations to a liquid layer of finite thickness. Instead of an analytic approach, we solve the Landau-Boltzmann equation by discretization and numerical inversion of the resulting large matrix.

There are a several motivations for the present work. First, in previous work the response of a Fermi liquid on a vibrating cylinder was studied \cite{VTletter,VTlong}. Such a calculation is computationally demanding and therefore it is of interest to study similar phenomena in the simpler planar geometry. Second, the behavior in a finite layer is much more diverse than in the thick layer limit. An indication of this is already given in the torsional oscillator calculations \cite{Virtanen12}. Third, the experiments by Casey et al \cite{Casey04} and Dimov et al \cite{Dimov10} show unexpected decoupling of the liquid  from the substrate. Previous analysis of these experiments neglected the Fermi-liquid interactions \cite{Sharma11}, and therefore we wanted to check if these have an effect. Fourth, we were interested to check if finite thickness effects could have affected previous experiments, and to predict the outcome of possible future experiments. Fifth, understanding the Fermi-liquid interactions in the normal state could form a useful step for properly incorporating the Fermi-liquid effects in the acoustic impedance of the superfluid state \cite{Nagai08}.

We use Fermi-liquid equations in the relaxation-time approximation and including interaction effects up to second order in spherical harmonics. The approach includes
the effect of the collective transverse zero-sound mode as well as incoherent quasiparticles.
In the limit of short mean free path of the quasiparticles, the flow of a Fermi liquid obeys Navier-Stokes equations. The leading correction to this hydrodynamic limit arises as ``slip'' in the boundary conditions \cite{HojgaardJensen80,Einzel97}. In our numerical solution no such expansion is made and  thus the slip effect is included in all orders. 

The calculations are mainly aimed for experiments in liquid \textsuperscript{3}He. However, the results apply also to simultaneous presence of a Bose condensate \cite{Khalatnikov69,TVperus}, and thus can also be applied to mixtures of \textsuperscript{3}He and \textsuperscript{4}He. Reviews of the acoustic impedance studies in both normal and superfluid \textsuperscript{3}He has been given by Halperin and Varoquaux \cite{Halperin90} and Okuda and Nomura \cite{Okuda12}. A theoretical review is given by Nagai et al\cite{Nagai08}. Recently experiments using a planar micromechanical oscillator in normal $^3$He have been made by Gonzalez et al and are analyzed using slip theory\cite{Gonzalez16}.

This paper is structured in the following way. In Sec.\ \ref{problem} we state the basic Fermi-liquid equations and transform them to a form suitable for numerical solution. In Sec.\ \ref{limit} we take a look at different limiting cases. In Sec.\ \ref{numeric} we explain the numerical method and finally in 
Sec.\ \ref{results} we present and comment the results.

\section{Fermi liquid equations} \label{problem}

\subsection{Equations of motion}

We study the linear response of a Fermi liquid film to the transverse oscillations of a planar substrate. We derive an expression for the  acoustic impedance $Z=F/u$, where $F$ is the force on the liquid per unit area of the film, and $u$ the velocity amplitude of the substrate, which is assumed to oscillate at angular frequency $\omega$. In this section we transform analytically the equations of a Fermi liquid to a  form that then can be solved numerically.  The transformation could be done by making only slight modification to the derivation by Flowers and Richardson~\cite{Flowers78}. Here we present a more general derivation that directly utilizes the distribution function defined in terms of  momentum direction and energy instead of momentum. The two distributions appear in several works\cite{Abrikosov59,Baym91,Bekarevich61,Flowers78}, and their relation is clearly pointed out in Ref.\ \onlinecite{Serene83}. We use the notation of a recent work that also includes the effect of condensed bosons \cite{TVperus}. We start by considering a pure fermion system, and postpone the minor effect of the bosons to Sec.\ \ref{s.bfl}.

Fermi liquid theory deals with quasiparticles with momenta $\bm p$ close to the Fermi surface, $p\approx p_F$. Assuming no spin dependence, the quasiparticle distribution function and energy can be written as
\begin{equation}
n_{\bm{p}}(\bm{r},t) = n_0(\epsilon_{\bm p}(\bm{r},t)) + \delta \bar n (\hat{\bm p},\epsilon_{\bm p}(\bm{r},t),\bm r,t),
\end{equation}
\begin{equation}
\epsilon_{\bm p}(\bm{r},t) = \epsilon^{(0)}_p + \delta \epsilon_{\hat{\bm p}}(\bm{r},t).
\end{equation}
Here $\hat{\bm p}=\bm p/p$ is the momentum direction, $n_0(\epsilon)=1/(e^{\epsilon/T}+1)$ the Fermi function, $\epsilon^{(0)}_p=v_{\rm F}(p-p_F)$ the unperturbed quasiparticle energy, $T$ the temperature, $v_F=p_{\rm F}/m^*$ the Fermi velocity and $m^*$ the effective mass. We also define the energy-integrated distribution function
\begin{eqnarray}
\psi_{\hat{\bm p}}(\bm r,t)=\int 
\delta \bar n (\hat{\bm p},\epsilon,\bm r,t)d\epsilon.
\label{e. psidef}\end{eqnarray}

In the relaxation-time approximation, the linearized kinetic equation can be written in a closed form for $\psi_{\hat{\bm p}}$. One gets the equations 
\begin{eqnarray}
\frac{\partial}{\partial t}(\psi_{\hat{\bm p}}-\delta\epsilon_{\hat{\bm p}})
+v_{\rm F}\hat{\bm p}\cdot\bm\nabla\psi_{\hat{\bm p}}=
-\frac{1}{\tau}\left[\psi_{\hat{\bm p}}-\langle\psi_{\hat{\bm p}'}\rangle_{\hat{\bm p}'}
-3\langle P_1(\hat{\bm p}\cdot\hat{\bm p}')\psi_{\hat{\bm p}'}\rangle_{\hat{\bm p}'}
+5(\xi_2-1)\langle P_2(\hat{\bm p}\cdot\hat{\bm p}')\psi_{\hat{\bm p}'}\rangle_{\hat{\bm p}'}\right],
\label{e.lbrel2c}\end{eqnarray}
\begin{eqnarray}
\delta\epsilon_{\hat{\bm p}}
=\sum_{l=0}^\infty
\frac{ F_l}{1+F_l/(2l+1)}
\langle P_l(\hat{\bm p}\cdot\hat{\bm p}')\psi_{\hat{\bm p}'}\rangle_{\hat{\bm p}'}.
\label{e.deltae2}\end{eqnarray}
Here $F_l$ with $l=0, 1,\ldots$ are the interaction parameters, $P_l$ the Legendre polynomials [$P_0(x)=1$, $P_1(x)=x$, $P_2(x)=\frac12(3x^2-1)$,\ldots] and $\langle \ldots\rangle_{\hat{\bm p}}$ the average over the unit sphere of momentum directions. In a pure fermion system $F_1$ is related to the ratio of effective and particle masses, $m^*/m=1+\frac13F_1$. Equations \eqref{e.lbrel2c}-\eqref{e.deltae2} are the same as derived in Ref.\ \onlinecite{TVperus} except the following generalization. We have allowed two relaxation times, $\tau_2=\tau/\xi_2$  for a quadrupole deformation of the Fermi surface and $\tau$ for all higher order deformations.  

In the following we neglect Fermi-liquid interaction coefficients beyond second order, $F_l=0$ for $l>2$. We also assume harmonic time dependence $\propto\exp(-i\omega t)$. These allow to write 
Eqs.\ \eqref{e.lbrel2c}-\eqref{e.deltae2} to the form
\begin{eqnarray}
\frac{\tau v_{\rm F}}{a}\hat{\bm p}\cdot\bm\nabla\psi_{\hat{\bm p}}+\psi_{\hat{\bm p}}
-\frac{ 1/a+F_0}{1+ F_0}\langle P_0(\hat{\bm p}\cdot\hat{\bm p}')\psi_{\hat{\bm p}'}\rangle_{\hat{\bm p}'}
-3b\langle P_1(\hat{\bm p}\cdot\hat{\bm p}')\psi_{\hat{\bm p}'}\rangle_{\hat{\bm p}'}
-c\langle P_2(\hat{\bm p}\cdot\hat{\bm p}')\psi_{\hat{\bm p}'}\rangle_{\hat{\bm p}'}=0.
\label{e.lbrel4f}\end{eqnarray}
We have defined dimensionless complex constants
\begin{align}
a&=1-i\omega\tau\label{aa},\\
 b &= \frac{1/a+F_1/3}{1+F_1/3}\label{ab},   \\ 
c &= \frac{5/a+F_2}{1+F_2/5} -\frac{5\xi_2}{a}\label{ac},
\end{align}
in accordance with Ref.\ \onlinecite{Flowers78}.

 We choose $z$ axis perpendicular to the liquid film and assume homogeneity in the $x$-$y$ plane.  The $x$ axis is chosen parallel to the oscillation of the wall. With these assumptions the most general distribution allowed in linear response can be written as
\begin{equation}
\psi_{\hat{\bm p}}(\bm r)=\hat p_x\psi(\hat p_z,\zeta).
\label{nkxf}\end{equation}
We also have defined $\zeta = z/d$ as the dimensionless $z$ coordinate, where $d$ is the thickness of the liquid film.
The form \eqref{nkxf} allows to simplify the averages
\begin{align}\langle P_0(\hat{\bm p}\cdot\hat{\bm p}')\psi_{\hat{\bm p}'}\rangle_{\hat{\bm p}'}
&=0,
\nonumber\\
\langle P_1(\hat{\bm p}\cdot\hat{\bm p}')\psi_{\hat{\bm p}'}\rangle_{\hat{\bm p}'}
&={\textstyle\frac{1}{4}} \hat p_xg_1(\zeta),
\nonumber\\
\langle P_2(\hat{\bm p}\cdot\hat{\bm p}')\psi_{\hat{\bm p}'}\rangle_{\hat{\bm p}'}
&={\textstyle\frac{3}{4}} \hat p_x\hat p_zg_2(\zeta).
\label{e.lbrelxz}\end{align}
Here the first average vanishes because transverse oscillations do not change the density of the liquid. The latter two averages depend on the integrals\cite{Flowers78}
\begin{subequations}
\label{phibar} 
\begin{align}
g_1(\zeta) &= \int_{-1}^{1}d\mu(1-\mu^2)\psi(\mu, \zeta) \label{phibar1}, \\
g_2(\zeta) &= \int_{-1}^{1}d\mu\,\mu(1-\mu^2)\psi(\mu, \zeta).\label{phibar2}
\end{align}
\end{subequations}
 Inserting these into the kinetic equation \eqref{e.lbrel4f}  gives
\begin{equation}\label{etabar}
\frac{\mu}{h}\frac{\partial}{\partial\zeta}\psi(\mu, \zeta) + \psi(\mu, \zeta) - {\textstyle\frac{3}{4}}bg_1(\zeta) - {\textstyle\frac{3}{4}}c\mu g_2(\zeta) = 0.
\end{equation}
We have abbreviated the equation by defining $\mu=\hat p_z$ and one more complex coefficient
\begin{equation}
h =\frac{a d}{v_F\tau} = \frac{d}{l\xi_2} - i\Omega(1+F_1/3).
\end{equation}
The latter form expresses $h$ using the dimensionless parameter
\begin{equation}
\Omega=\frac{\omega d}{v_F(1+F_1/3)}
\label{Omegae}\end{equation}
and the mean free path $l$. Since $\tau_2$ is the effective relaxation time in the hydrodynamic limit,  the quasiparticle mean free path is defined as $l=v_F\tau_2=v_F \tau/\xi_2$. A convenient set of dimensionless parameters that define the problem is formed by $\Omega$, $l/d$, $\xi_2$, $F_1$ and $F_2$.

We may further solve for $\psi$ by integrating the kinetic equation \eqref{etabar} from $\zeta_0$ to $\zeta$:
\begin{equation}\label{etrans}
 \psi(\mu,\zeta) = \psi(\mu,\zeta_0)e^{\frac{h}{\mu}(\zeta_0-\zeta)} 
+ \frac{3}{4}\frac{h}{\mu}\int_{\zeta_0}^\zeta e^{\frac{h}{\mu}(\zeta'-\zeta)}[bg_1(\zeta') + c\mu g_2(\zeta')]d\zeta'.
\end{equation}
We use this equation to integrate in the direction of particle propagation. That is, we integrate in the direction of increasing $\zeta$ for angles pointing up ($\mu>0$) and decreasing $\zeta$ for angles pointing down ($\mu<0$), see Fig.\ \ref{fig:FLtrajec}.

In addition to the equations of motion, we need to specify boundary conditions.  
We set a stationary plane at $z=d$, that is $\zeta=1$. We assume that the quasiparticle scattering at this surface is diffusive except for a fraction $s_2$ of quasiparticles, which is scattered specularly. The case $s_2=1$ then mimics a free surface of a liquid. We set the oscillating wall at $z=\zeta=0$. Its velocity is $u\hat{\bm x}e^{-i\omega t}$. We assume that the quasiparticle scattering at this wall is diffusive except for a fraction $s_1$ of quasiparticles, which is scattered specularly.  These imply the following boundary conditions for the distribution function \cite{TVperus}
\begin{subequations}
\label{e.bcon} 
\begin{align}
\psi(\mu < 0,  1) &=s_2\psi(-\mu, 1)\label{bcus},\\
\psi(\mu > 0,  0) &=s_1\psi(-\mu,  0) + (1-s_1)p_F u.\label{bcls} \end{align}
\end{subequations}

Note that because of symmetry, it is also possible to consider the liquid between two equally oscillating walls by setting $s_2=1$ and $2d$ being the distance between the walls.

We see from the boundary conditions \eqref{e.bcon} that the distribution function has to be proportional to $p_F u$. We can then factor out $p_F u$ for the sake of numerical convenience by defining the effective fields
\begin{equation}\label{nubar}
\psi^e=\frac{\psi}{p_F u},\quad g_1^e=\frac{g_1}{p_F u},
\quad g_2^e=\frac{g_2}{p_F u}.
\end{equation}
This simplifies the boundary conditions \eqref{e.bcon} but the bulk equations \eqref{phibar} and \eqref{etrans} remain the same for the effective fields. Equations  \eqref{phibar},  \eqref{etrans} and \eqref{e.bcon} constitute the set of integral equations and boundary conditions that we can solve numerically.

\subsection{Observables}

The macroscopic forces acting in the liquid are obtained by calculating the stress tensor
\begin{equation}
\Pi_{ij}=3n\langle \hat{p}_i\hat{p}_j\psi_{\hat{\bm p}}\rangle_{\hat{\bm p}},\end{equation}
where $n=p_F^3/3\pi^2\hbar^3$ is the number density of the fermions.
The shear force, the $xz$ component of the stress tensor, can be evaluated using \eqref{nkxf} and \eqref{phibar2}, which gives
\begin{equation}
\Pi_{xz}(\zeta) =  \frac{ 3}{4} ng_2(\zeta) = \frac{ 3}{4} np_Fug_2^e(\zeta).
\label{stretesz}\end{equation}
The acoustic impedance of the liquid film is then 
\begin{equation}
Z = \frac{\Pi_{xz}(\zeta=0)}{u}= \frac{ 3}{4} np_Fg_2^e(\zeta=0).
\label{zgephi20}\end{equation}

The mass current in the liquid is
\begin{equation}
\bm J=\frac{mp_F^2}{\pi^2\hbar^3}\langle \hat{\bm p}\psi_{\hat{\bm p}}\rangle_{\hat{\bm p}}.\end{equation}
Evaluating this using \eqref{nkxf} and \eqref{phibar1} gives that the current is in the $x$ direction and its magnitude
\begin{eqnarray}\label{velo}
J(\zeta)=\frac{mp_F^2}{4\pi^2\hbar^3}g_1(\zeta)=\frac34\rho ug_1^{\rm e}(\zeta),
\end{eqnarray}
where the liquid density $\rho=mn$.
Thus $\frac34g_1^{\rm e}(\zeta)$ can be interpreted as the average velocity normalized by the substrate velocity $u$. In the hydrodynamic limit this should approach unity at the substrate ($\zeta=0$). This is the velocity field in the transverse wave and should not be confused with the velocity of the wave itself.

\subsection{Bose-Fermi liquid}\label{s.bfl}

The theory above can straightforwardly be generalized to the simultaneous presence of condensed bosons \cite{Khalatnikov69,TVperus}. In transverse oscillations the superfluid component remains at rest. In the notation of Ref.\ \onlinecite{TVperus}, $\bm v_s=0$ and $\delta \mu_{B}=0$. 
The equations, in particular Eqs.\ (34) and (35), of Ref.\ \onlinecite{TVperus} reduce to those ones in the present paper. The only difference is that Eq.\ \eqref{velo} gives only the fermionic contribution to the current. In order to get the total mass current, the fermion density $\rho$ should be replaced by the normal fluid density $\rho_n=m^*n/(1+F_1/3)$.

\section{Limiting cases} \label{limit}

There are limiting cases that are worth of studying separately.  In some cases analytic solutions are known.

\subsection{Hydrodynamic limit}

At high temperatures quasiparticle collisions become frequent and thus the mean free path is short. In the hydrodynamic regime $l$ is short compared to other length scales, $l\ll d$ and $l\ll v_F/\omega$. In this regime the Fermi-liquid theory implies equations of motion that are the well known hydrodynamic, or Navier-Stokes equations. Solving these for laminar flow between two parallel planes is a common exercise in books on hydrodynamics \cite{LLf,Batchelor67}. We need to consider two boundary conditions for the top surface; a free liquid surface ($s_2=1$) and an unmoving solid surface, i.e.\ Couette flow ($s_2=0$). For the oscillating wall, we assume no slip ($s_1=0$). We use the Navier-Stokes equation to calculate the force at the boundary of the liquid and the oscillating surface. This way one gets the acoustic impedance
\begin{equation}
Z = \frac{F}{u} = \frac{\rho\omega\delta}{2} (1 - i)\frac{e^{2(1 - i)d/\delta}\mp 1}{e^{2(1 - i)d/\delta}\pm 1}.
\label{e.zhydro}\end{equation}
The upper signs stand for a free liquid surface and the lower signs for Couette flow. These are identical when $d/\delta$ is large. Here $\delta$ is the viscous penetration depth, which is related to other parameters by equations
\begin{equation}
\delta = d\sqrt{\frac{2 l}{5 \Omega d}} = \sqrt{\frac{2{v_F}^2 \tau_2(1+F_1/3)}{5\omega}}.
\label{e.pentdepth}\end{equation}
We see that the essential dimensionless parameter in \eqref{e.zhydro} is $\delta/d$.

\subsection{Ballistic limit}

At very low temperatures quasiparticle collisions become so infrequent that they may be neglected. This is the ballistic regime, where the quasiparticles travel between the oscillating plane and the liquid surface without encountering each other. In the ballistic limit we take $l\rightarrow\infty$. In the general case, this does not lead to any simplification of the kinetic equation \eqref{etrans}, as the limit $a\rightarrow \infty$ still leaves $b$  (\ref{ab}) and $c$  (\ref{ac}) finite. However, if  we also set the Fermi liquid interactions $F_1$ and $F_2$ to zero, then $b=c=0$. We call this the {\em ballistic gas limit}. This leads to essential simplification of the kinetic equation
\eqref{etrans}, which reduces  to the form
\begin{equation}
\psi^e(\mu,\zeta) = e^{\frac{h}{\mu}(\zeta_0-\zeta)} \psi^e(\mu,\zeta_0).
\end{equation}
By using this equation and the boundary conditions \eqref{e.bcon} to traverse rectangular paths in the $(\mu,\zeta)$ space (Fig.\ \ref{fig:FLtrajec}), we obtain the quasiparticle distribution
\begin{alignat}{2}
\mu &> 0: \quad &\psi^e(\mu,0) &= \frac{1-s_1}{1-s_1 s_2 e^{-2h/\mu}}, \\	
\mu &< 0:  &\psi^e(\mu,0) &= \frac{s_2(1-s_1) e^{2h/\mu}}{1-s_1 s_2 e^{2h/\mu}}.
\end{alignat}
We can now compute $g_2^e$ at the oscillating wall, which gives the acoustic impedance 
\begin{equation}
Z = \frac34np_F(1-s_1)\int_{0}^{1}d\mu\,\mu(1-\mu^2)\frac{1 - s_2 e^{2i\Omega/\mu}}{1 - s_1 s_2 e^{2i\Omega/\mu}}.
\label{zballim}\end{equation}
We see that this depends essentially on $\Omega$ \eqref{Omegae}. Note that this result is valid only in the case of  $F_1=F_2=0$.

\subsection{Thick film limit}\label{s.tfl}

Let us consider the case of a very thick film, $d\rightarrow\infty$. This was first studied by Bekarevich and Khalatnikov \cite{Bekarevich61} and more generally by Flowers and Richardson \cite{Flowers78}. They found an analytic solution. Since the result is not simple, we will not reproduce it here. Instead we point out that there are two variational ansatz solutions given in equations (3.32) and (3.38) of Ref.\ \onlinecite{Richardson78}. The essential dimensionless parameter in the thick film limit is  $\Omega l/d=\omega\tau_2/(1+\frac13F_1)$.

\section{Numerical Solution} \label{numeric}

The search for the numerical solution begins with the discretization of the $(\mu,\zeta)$ space. Generally the $\zeta$ axis is divided into segments of equal length so that $\zeta_j=(j-1)/(n-1)$, where $j=1,...,n$. For high temperatures, the wave emanating from the oscillating wall will not penetrate deep into the liquid layer. In this case, instead of equally spaced lattice points, it is more efficient to use a discretization that places lattice points more densely in the vicinity of the oscillating wall. 

Different discretization schemes may be employed for the $\mu$ axis. We approximate integrals over $\mu$ \eqref{phibar} with
\begin{equation}
\int_{-1}^1 f(\mu)d\mu \approx \sum_{i=1}^m w_i f(\mu_i).
\end{equation}
The values $\mu_i$ and the weights $w_i$ are selected using Gaussian quadrature. We use an even number $m$ of $\mu_i$ values in order to avoid the $\mu=0$ point, which could require special treatment.

The integration over $\zeta$ \eqref{etrans} is made only between neighboring discretized points. Instead of a simple trapezoidal formula we use
\begin{equation}
\begin{split}
\int_{\zeta_0}^{\zeta} e^{\alpha \zeta'} f(\zeta')d\zeta' \approx w_1 f(\zeta_0) + w_2 f(\zeta), \\
w_1=\frac{1}{\alpha^2(\zeta-\zeta_0)}[e^{\alpha\zeta}-(1+\alpha(\zeta-\zeta_0))e^{\alpha\zeta_0}], \\
w_2=\frac{1}{\alpha^2(\zeta-\zeta_0)}[e^{\alpha\zeta_0}-(1-\alpha(\zeta-\zeta_0))e^{\alpha\zeta}],
\end{split}
\end{equation}
where $\alpha=h/\mu$. This method allows better accuracy if the exponential factor inside the integral in equation \eqref{etrans} varies rapidly. 

The discrete versions of the integral equations \eqref{phibar}, \eqref{etrans}  and \eqref{e.bcon} provide a network of linear dependencies between $\psi(\mu_i,\zeta_j)$, $g_1(\zeta_j)$ and $g_2(\zeta_j)$. We now form a vector $\Psi$ that holds all these variables in the following fashion:
\begin{equation}
\Psi = (\psi^e(\mu_1,\zeta_1),...,\psi^e(\mu_m,\nu_n);
g^e_1(\zeta_1),...,g^e_1(\zeta_n);g^e_2(\zeta_1),...,g^e_2(\zeta_n)).
\end{equation}
The length of this vector is $d=mn+2n$. The network of linear dependencies may now be represented in the form
\begin{equation}
\Psi = D\Psi + B.
\label{deqmat}\end{equation}
Here the left hand side represents the left hand side of equations \eqref{phibar}, \eqref{etrans}  and \eqref{e.bcon}, and correspondingly for the right hand sides.  The matrix $D$ is of dimension $d\times d$. The vector $B$ is the inhomogeneity term arising from the non-specular scattering at the oscillating wall in the last term of 
Eq.\ \eqref{bcls}. Equation \eqref{deqmat} is a system of linear equations for $\Psi$, whose solution can be written as
\begin{equation}
\Psi = (I-D)^{-1}B.
\label{deqmats}\end{equation}

The task is now to first assemble the $D$ matrix and then solve the linear system \eqref{deqmat} to get $\Psi$.  The acoustic impedance \eqref{zgephi20} is then obtained by picking out the element that corresponds to $g_2^e(0)$. 
We can also pick out $g_2^e$ at any $\zeta$ to get the stress tensor \eqref{stretesz} within the liquid or  $g_1^e$ to get the transverse velocity field \eqref{velo}. While the dimension $d\times d$ may be very large, the $D$ matrix only has $7mn-4m$ elements that can be non-zero. We use sparse matrix methods for solving the the inverse \eqref{deqmat}. This requires specifying those elements of the matrix $D$ that can be nonzero so that the zero elements never need to be addressed. Since $B$ is sparse and only a few elements of  $\Psi$ are of interest, there is no need to calculate the whole inverse matrix $(I-D)^{-1}$.

\section{Results} \label{results}

Before presenting the results of the numerical calculations,  we outline the parameter values that define experimentally relevant conditions. Foremost there are the Fermi-liquid interaction parameters $F_1$ and $F_2$ that describe the forces between the quasiparticles. In pure $^3$He the parameter $F_1$ is pressure dependent and its value has been determined experimentally \cite{Greywall86}. Some notable values are $F_1=5.4$ at zero pressure and $F_1=13.3$ at the melting pressure.  There are no generally accepted values for $F_2$, but it is thought to range between $-1$ and $1$ \cite{Engel85,Nettleton78,Halperin90}. A requirement for the existence of transverse zero sound is expected to be  \cite{Fomin68,Lea73,Fomin76,Flowers78,Halperin90}
\begin{equation}
F_1+\frac{3F_2}{1+\frac15F_2}>6
\label{e.etrzs}\end{equation}
in our model, where $F_n$ with $n>2$ are neglected.
In addition, we have the ratio $\xi_2= \tau/\tau_2$ of the two relaxation times, for which the value $\xi_2= 0.35$ has been suggested \cite{Flowers78}. Other dimensionless parameters are $\Omega=\omega d/v_F(1+\frac13F_1)$ \eqref{Omegae} and $l/d$. The former depends essentially on the film thickness $d$ and frequency $\omega$ whereas the latter depends essentially on the temperature as the mean free path $l\propto T^{-2}$ in the Fermi liquid regime. We use the coefficient $lT^2$ based on viscosity measurements as given in Tables III and IV of Ref.\ \onlinecite{Wheatley75} and $m^*/m$ from Ref.\ \onlinecite{Greywall86}. The specularity parameters $s_1$ and $s_2$ define conditions at the two liquid boundaries.

We show plots of transverse acoustic impedance $Z=Z'+iZ''$. The real part of the impedance corresponds to dissipation and the imaginary part to reactance. We display parametric plots of $Z$ as well as separate plots of $Z'$ and $Z''$ as functions temperature or $l/d$, for which we use a logarithmic scale. 

\begin{figure*}[bt]
        \centering
                \includegraphics[width=0.8\textwidth]{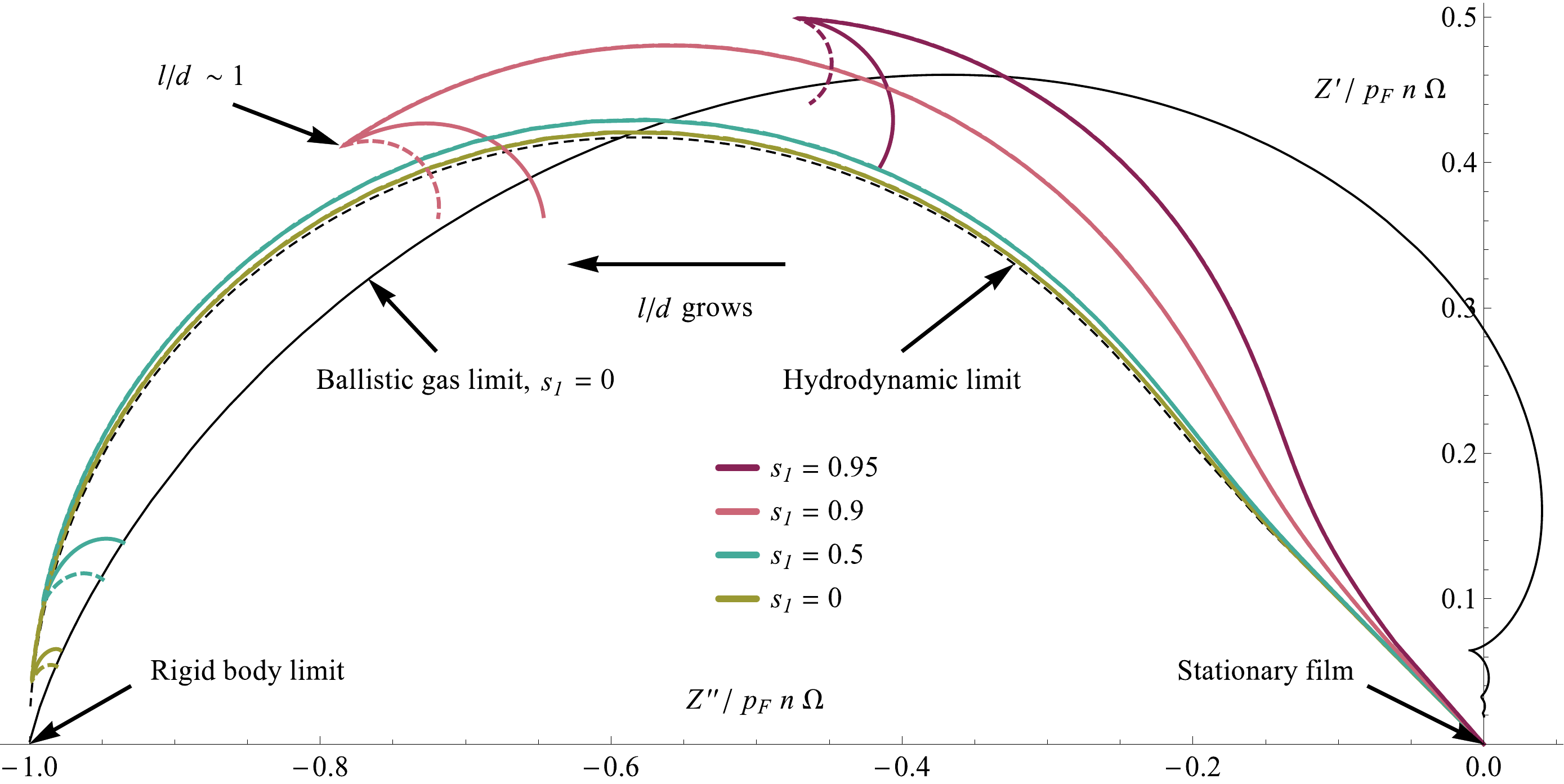}
        \caption{Parametric plot of $Z=Z'+iZ''$ with $l/d$ as a variable for a small $\Omega = 10^{-3}$ \eqref{Omegae}. Different curves correspond to different values of the Fermi-liquid parameter $F_1$ and the specularity $s_1$. The liquid has a free top surface ($s_2=1$) and the parameter $s_1$ controls the specularity of the oscillating bottom wall. The  arrows point out the direction of increasing $l/d$, the various analytical limiting cases and the approximate point where $l/d\sim 1$. For solid curves $F_1=0$ and for dashed curves $F_1=5.4$. The hydrodynamic limit \eqref{e.zhydro} is shown by black dashed line. The ballistic gas limit \eqref{zballim}, represented by the black solid line, was computed using $s_1=0$ and $F_1=0$. Other parameters are $F_2=0$ and $\xi_2=1$.}\label{fig:tinyomega}
\end{figure*}

We consider first the case of small $\Omega$ \eqref{Omegae}, $\Omega\ll 1$.
Supposing there are waves whose speed is on the order of the Fermi velocity $v_F$, the condition $\Omega\ll 1$ means that the film thickness is much smaller than the wave length of such waves. That is, there is flow but no space for propagating waves. For small $\Omega$ it is convenient to scale the impedance by $\Omega$. This produces pictures like Fig.\ \ref{fig:tinyomega}. With this scaling the hydrodynamic limit curves  \eqref{e.zhydro} at different $\Omega$ coalesce into a single curve, which is depicted here by the black dashed line. 

Let us analyze a curve in Fig.\ \ref{fig:tinyomega} in the order of increasing $l/d$. 
The origin, $Z=0$, corresponds to the stationary film limit, where the liquid remains at rest in spite of the oscillation of the substrate. For small but finite $l$ the diffusive waves generated at the oscillating surface penetrate to depth
$\delta$  \eqref{e.pentdepth} into the liquid. These give rise to $Z$ on a straight line in the direction $1-i$ \eqref{e.zhydro}. The penetration depth increases with growing $l$. When $\delta\sim d$  the wave starts to feel the liquid surface, and the path in the $Z$ plane starts to curve.
For small $s_1$ the curves still stay close to the hydrodynamic limit for some range of $l$. The hydrodynamic limit curve \eqref{e.zhydro} continues towards the point $Z/p_Fn\Omega=-i$, which corresponds to rigid body motion of the liquid with the substrate.
Before reaching this point, the curves develop a cusp at $l\approx d$. With further increase of $l$ the system enters the ballistic, or Knudsen, regime. With $l\rightarrow \infty$ they reach the ballistic limit point. A set of ballistic limit points \eqref{zballim} in the noninteracting, $s_1=0$ case is shown by the black solid line in Fig.\ \ref{fig:tinyomega}.

We see that with the scaling of Figure \ref{fig:tinyomega}, the effect of the interaction parameter $F_1$ is limited to mean free paths $l>d$. The same holds for $F_2$ as well. Also, the effect of the interaction parameters is an order of magnitude smaller than the whole scale of $Z''$ in the figure. With increasing specularity of the oscillating wall, the film becomes more decoupled. However, rather high specularity ($>0.9$) is required to have a half of the liquid film mass decoupled in the ballistic limit and there still remains strong dissipation  that shows no sign of decreasing with increasing specularity.

The results above may be applied to the experiments by Casey et al\cite{Casey04} and Dimov et al\cite{Dimov10}, which report decoupling of the liquid  from the substrate with decreasing temperature. Our motivation was to check if Fermi-liquid interactions could be responsible for the decoupling. Since we see only minor decoupling, we have to conclude that our Fermi-liquid model is not capable to explain these experimental observations. 

\begin{figure*}[tb]
        \centering
                \includegraphics[width=\textwidth]{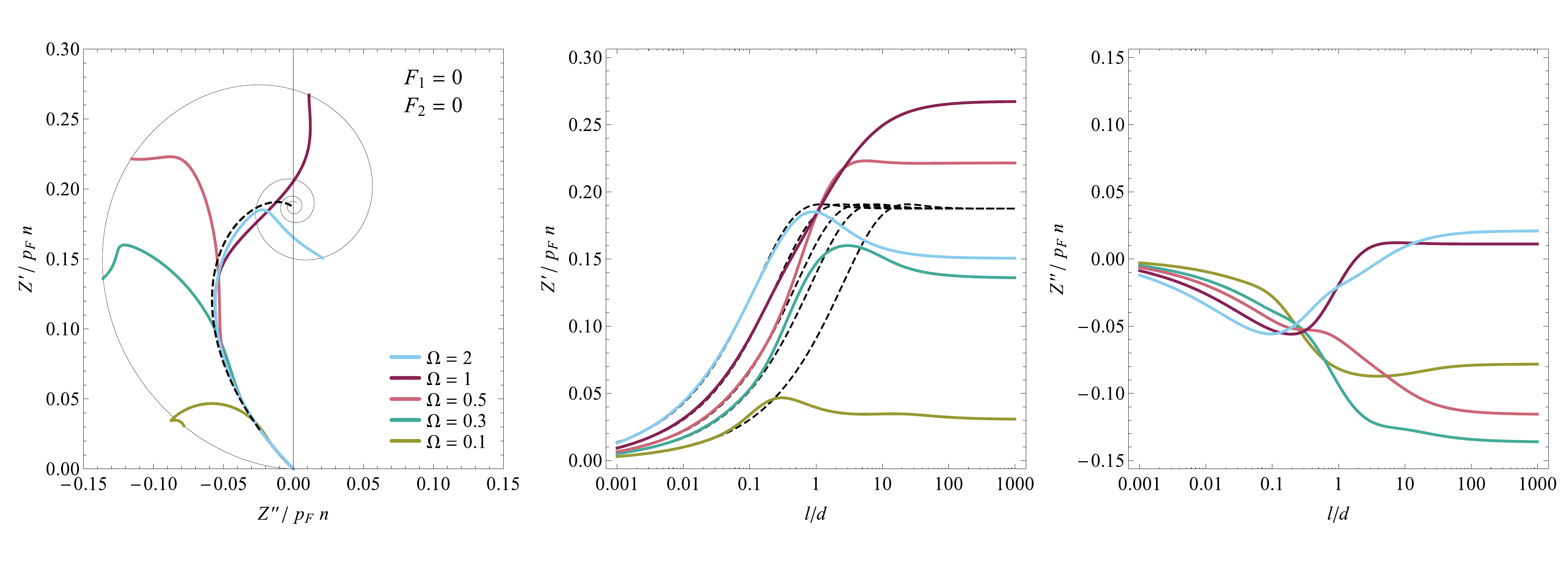}
        \caption{Acoustic impedance in the absence of Fermi-liquid interactions at different $\Omega$ \eqref{Omegae}. Displayed here are, from left to right, a parametric plot of the impedance and the real and the imaginary parts of the impedance as functions of $l/d$. Similar to Fig.\ \ref{fig:tinyomega}, the black solid curve represents the ballistic gas limit \eqref{zballim}. The dashed black curves represent the thick film limit according to Ref.\ \onlinecite{Richardson78}. In the parametric plot (left panel) the dashed lines collapse to a single curve but in the real part vs. $l/d$ (center panel) they are shifted from each other. For clarity, the thick film limit is omitted in the imaginary part vs. $l/d$ (right panel). The results for the finite film differ from the thick film limit when the liquid surface is felt. From the middle panel we see that this is the case when $l/d>0.1$. Other parameters are $s_1=0$, $s_2=1$, and $\xi_2= 0.35$.}\label{fig:nonscaled00}
\end{figure*}

Let us next consider the case of large $\Omega$. In this case, when waves develop, there is room for several wave lengths in the film. In this case the rigid body limit cannot be reached, and it is more convenient to analyze $Z$ without scaling with $\Omega$. An example of curves up to $\Omega=2$ is shown in Fig.\ \ref{fig:nonscaled00}.
We again follow one curve in the order of increasing $l/d$. Initially the curve starts from the origin along a straight line in the direction $Z''=-Z'$, similarly as in the case of small $\Omega$. However, for large $\Omega$ we exit the hydrodynamic regime before reaching $\delta\sim d$. This happens because $\omega\tau\sim\Omega l/d$ approaches unity. Thus, the impedance deviates from the hydrodynamic limit and follows the curve calculated in the thick film limit (Sec. \ref{s.tfl}). This curve is shown by dashed lines in Fig.\ \ref{fig:nonscaled00}. This continues as long as the waves generated at the oscillating wall start to feel the surface and are reflected back. At this point the curve deviates from the dashed line, as will be analyzed shortly. 

Figure \ref{fig:nonscaled00} depicts the special case that the Fermi liquid interactions vanish, $F_1=F_2=0$. This case is not realized in pure $^3$He. Experimentally a close case could be studied in mixtures of $^3$He and $^4$He, where the Fermi liquid interactions are weaker than in pure $^3$He [Ref.\ \onlinecite{VTlong}].  Interestingly, no waves are expected according to criterion \eqref{e.etrzs}. Still we see waves, the end points of the curves lie on the spiral, not at the end point of the dashed line. The spiraling down indicates damping of these waves. Note that the ballistic limit curves (black solid lines) in 
Figs.\ \ref{fig:tinyomega} and \ref{fig:nonscaled00} are the same, the difference comes only from the different scalings used.

We note that we did not succeed to compute numerically the exact analytic result of the thick film limit \cite{Flowers78} in the case of 
Fig.\ \ref{fig:nonscaled00}. This apparently has to do with some numerical problem when the inequality \eqref{e.etrzs} is not satisfied. Instead, we use the simpler of the approximate formulas, Eq.\ (3.32), of Ref.\ \onlinecite{Richardson78}. We see that the curves initially follow the thick film behavior, until the effect of the liquid surface appears. Comparing the real parts of the two solutions gives that this takes place around $l/d\sim 0.1$.

The center and right hand panels of Fig.{ \ref{fig:nonscaled00} show the real and imaginary parts of $Z$ plotted separately as functions of $l/d$. We see that both the real and imaginary parts of the impedance fully plateau as $l/d\to 1000$. This indicates that in the parametric plot the curves have arrived at static endpoints.

Figure \ref{fig:nonscaledF1} is similar to the previous one, but this time the first Fermi liquid parameter is set to $F_1=13.3$ and so the curves no longer end on the ballistic gas limit \eqref{zballim}. In this case the zero-sound criterion \eqref{e.etrzs} is satisfied and allows us to compute the exact thick-film solution, represented by the  black dashed line \cite{Flowers78}. Similar to the noninteracting case, the curves initially follow the thick film behavior, until the effect of the liquid surface appears. 

\begin{figure*}[tb]
        \centering
                \includegraphics[width=\textwidth]{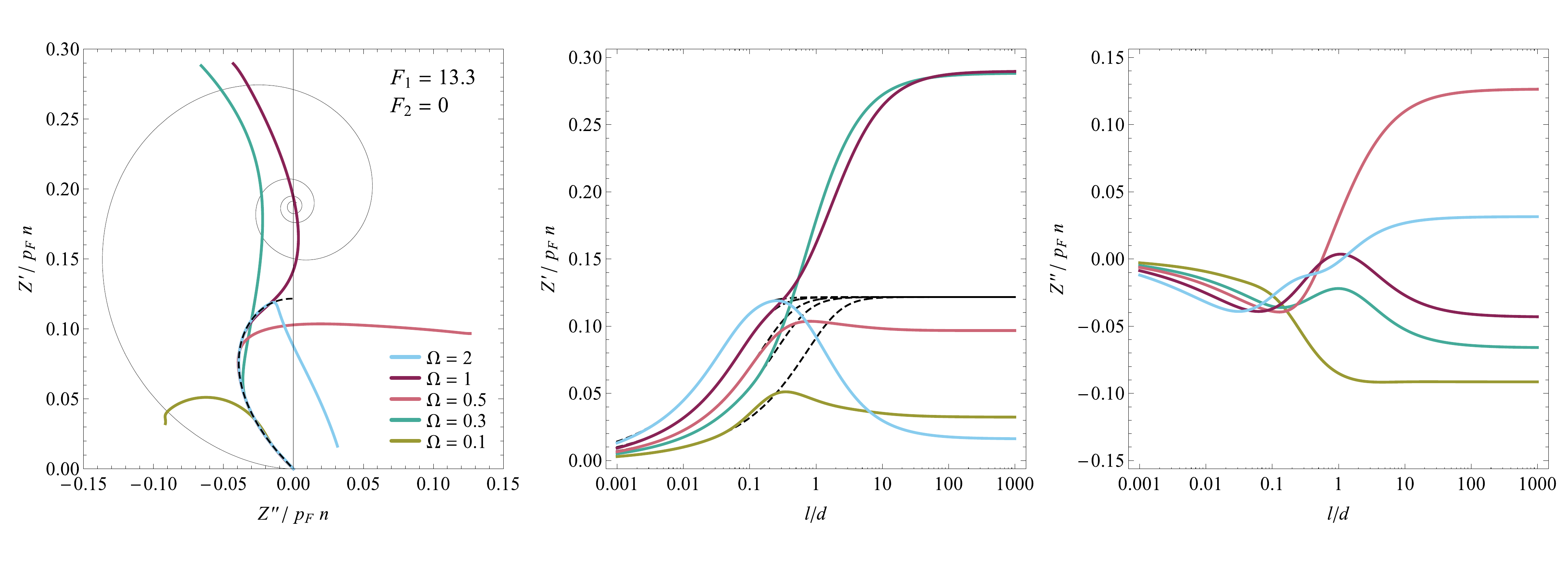}
        \caption{The same as Fig.\ \ref{fig:nonscaled00} except that $F_1=13.3$.  The ballistic gas line is the same as in Fig.\ \ref{fig:nonscaled00} to allow easier comparison of the figures. Because of interactions, the curves do not end on this line. The dashed black curves represent the exact thick film limit\cite{Flowers78}. Other parameters are $F_2=0$, $s_1=0$, $s_2=1$, and $\xi_2= 0.35$.}\label{fig:nonscaledF1}
\end{figure*}

Figure \ref{fig:nonscaledF2} is again similar to the previous two, but this time also the second interaction parameter has a nonzero value, $F_2=1$. Comparing this to Fig.\ \ref{fig:nonscaledF1}, we see that the effect of $F_2$ strongly increases for increasing $\Omega$.

\begin{figure*}[tb]
        \centering
                \includegraphics[width=\textwidth]{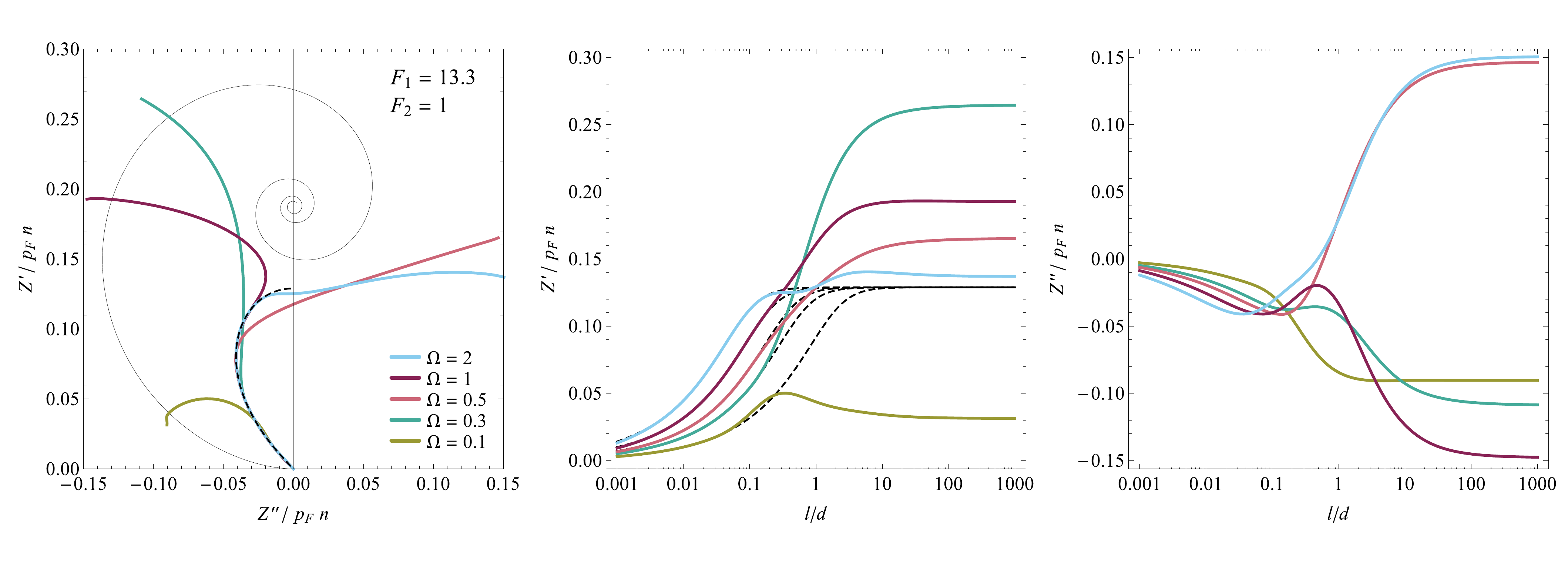}
        \caption{The same as Fig.\ \ref{fig:nonscaledF1} except that $F_2=1$. There is significant difference between the figures at large $\Omega$. }\label{fig:nonscaledF2}
\end{figure*}

We can understand the behavior of the ends of the curves in figures \ref{fig:nonscaledF1} and \ref{fig:nonscaledF2} as follows. With the condition \eqref{e.etrzs} satisfied, the damping of the waves is weak. This means that with increasing $\Omega$ the end points nearly circle around a point in the complex $Z$ plane without any apparent damping. The waves emanating from the oscillating wall are reflected back from the top surface. By changing the layer thickness, sound velocity or the oscillation frequency, we potentially alter the phase at which the waves return back to the oscillating surface. If the returning wave is in-phase with the wall oscillations, then the oscillations are amplified. Conversely, a returning wave in opposite phase leads to destructive interference. Changing the liquid layer thickness by a quarter of the wavelength results in the opposite phase and a deviation in the opposite direction from the thick film limit. An implication of this is that in order to see finite thickness effects, the boundaries of the liquid need to be accurately parallel.

In Fig.\ \ref{fig:RK} the layer thickness is fixed and the mean free path changes as a function of temperature. We have used parameter values that correspond to the experiment by Roach and Ketterson \cite{Roach76}. In the calculation the liquid is confined between two diffusely reflecting plates spaced $d=25\ \mu$m apart. We see that in this setting the presence of the top plate, which is seen as the bifurcation of the three curves, is only felt at temperatures below the superfluid transition temperature $T_c$. The liquid layer is simply too thick for the transverse sound wave to penetrate all the way to the other wall and back at temperatures above $T_c$. This can be confirmed in Fig.\ \ref{fig:velocity}, where the transverse velocity field \eqref{velo} is plotted as a function of temperature. 

\begin{figure*}[tb]
        \centering
                \includegraphics[width=\textwidth]{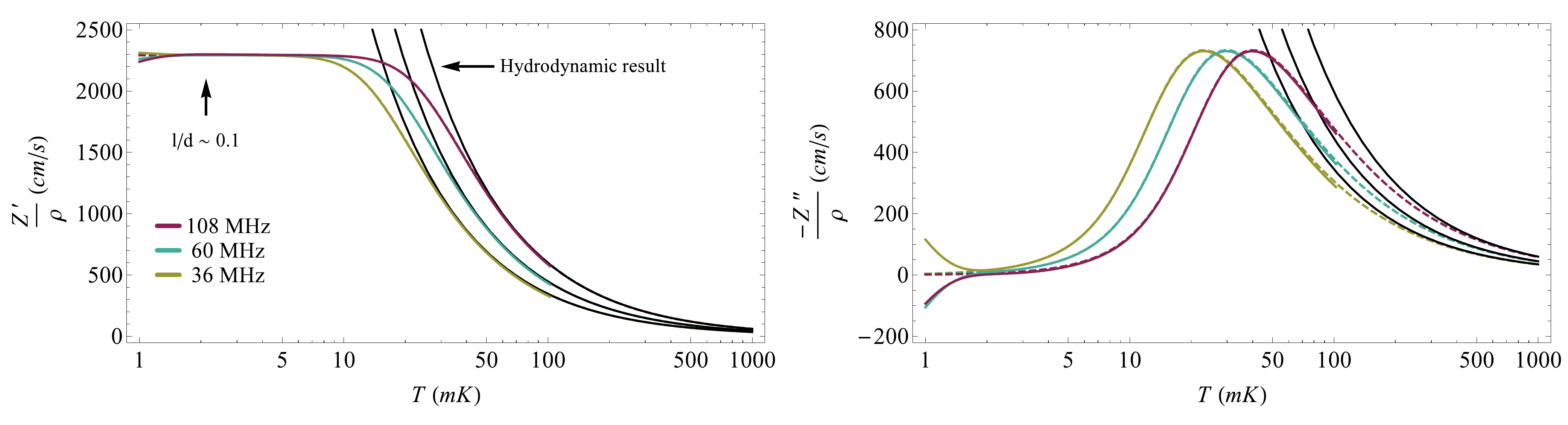}
        \caption{Acoustic impedance as a function of temperature for three different frequencies at the pressure of $23$ bar.  The solid curves are for a liquid confined between two diffusely reflecting plates spaced $25\ \mu$m apart, and correspond to $\Omega$ in the range from  31 to 93. The dashed curves give the thick film limit. The two cases differ only at low temperatures, the slight difference at high temperatures is due to numerical error. The black solid lines give the hydrodynamic solution \eqref{e.zhydro}. 
The vertical arrow denotes the superfluid transition point, $T_c=2.3$ mK. The parameters are $F_1=11.8$, $F_2=0$, $s_1=0$, $s_2=0$, and $\xi_2= 0.35$.}\label{fig:RK}
\end{figure*}

\begin{figure}[tb]
  \centering
    \includegraphics[width=0.7\columnwidth]{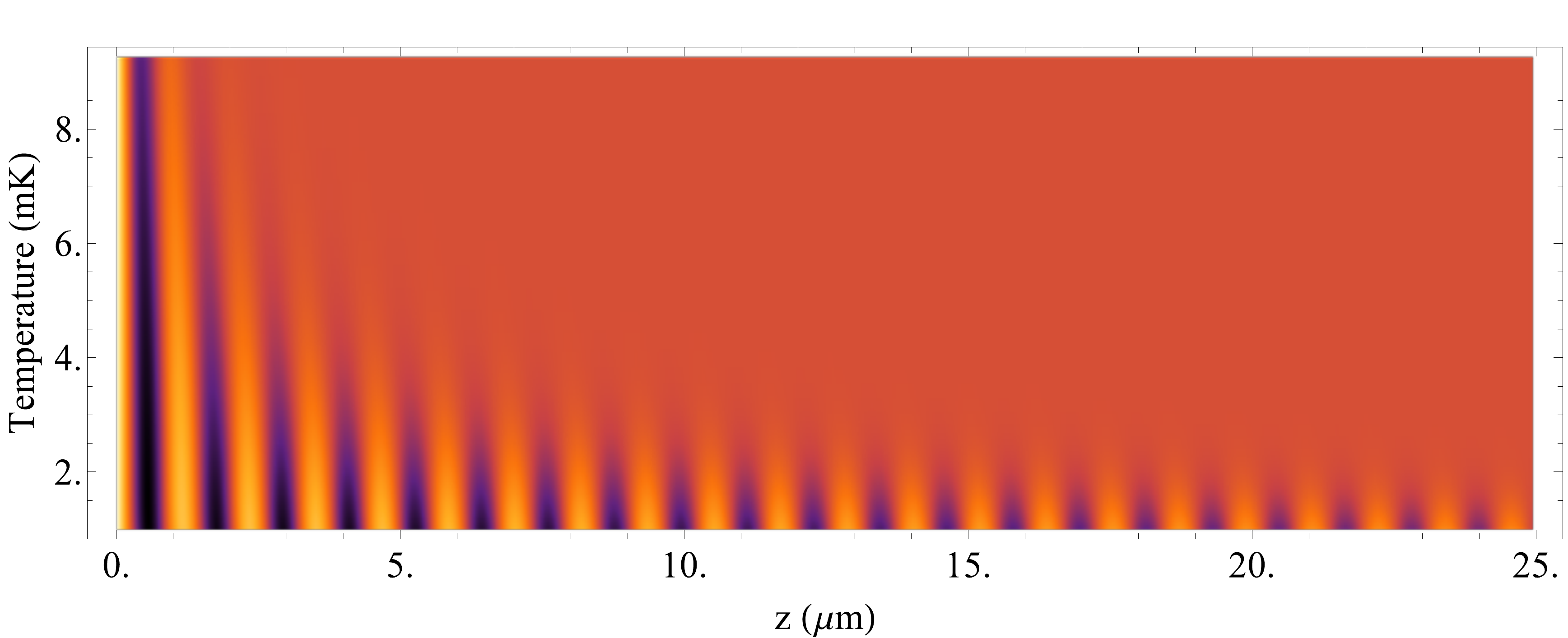}
  \caption{Transverse velocity field as a function of temperature and the distance from the oscillating wall. The transverse sound waves penetrate to the other wall at around one millikelvin. The parameters are the same as in Fig.\ \ref{fig:RK} corresponding to $36$ MHz. }
  \label{fig:velocity}
\end{figure}

The obvious thing to do is to repeat the computation in Fig.\ \ref{fig:RK} for a thinner film. By selecting $d = 2.5\ \mu$m and using the smallest oscillation frequency $36$ MHz, we conveniently have $\Omega\approx 3$ which, based on our previous analysis, is in the range where we should see large sensitivity to $F_2$. 
The results are shown in Fig.\  \ref{fig:F2row}. We have used two different values of $F_2$. Both the thick film and finite film solutions show sensitivity to $F_2$. The bifurcation between these two solutions happens well above the superfluid transition temperature. For both solutions an increase in $F_2$ results in an initially identical shift in the impedance but, in addition to this, the thin film solution is influenced by the top plate once the temperature gets sufficiently low. This is especially apparent for $Z''$, for which the thick film solutions converge as $T\rightarrow 0$.
\begin{figure*}[tb]
        \centering
                \includegraphics[width=\textwidth]{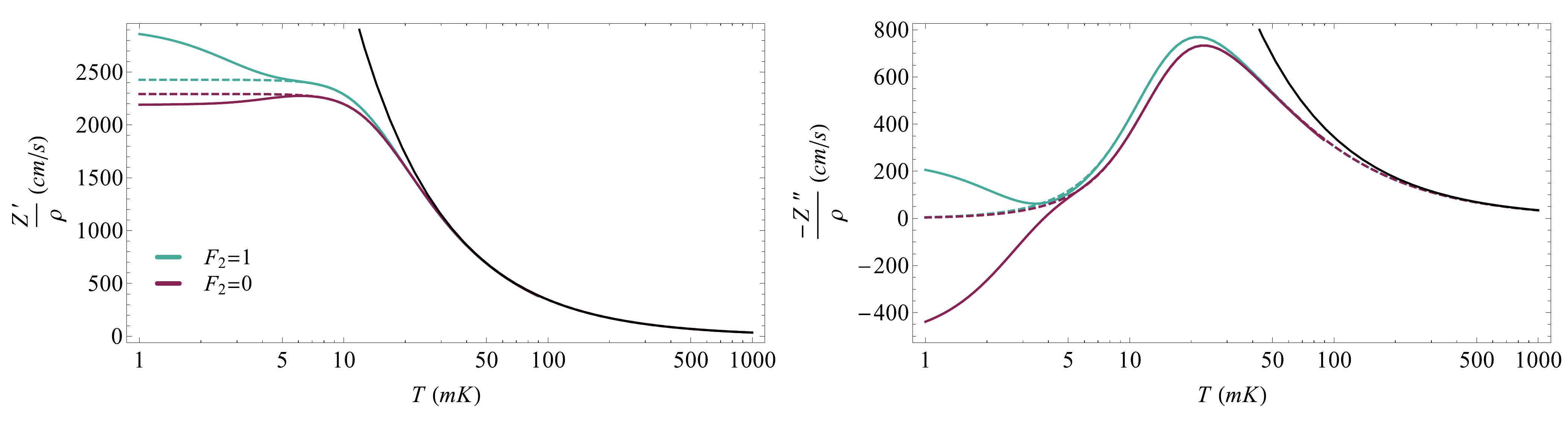}
        \caption{The effect of $F_2$ for a film of thickness $d = 2.5\ \mu$m. Other parameter values are pressure $23$ bar, frequency $36$ MHz, $F_1=11.8$, $s_1=s_2=0$, and $\xi_2=0.35$. Solid curves depict the finite film solution, dashed curves the thick film limit. The black solid line represents the hydrodynamic solution.}\label{fig:F2row}
\end{figure*}

\section{Summary}

We have formulated how to calculate the transverse acoustic impedance of a Fermi-liquid film. We have built up a  scheme for numerical evaluation. Some example results are presented in this paper aiming to clarify the case of a few known experiments and stimulate new ones. In the future we plan to extend the calculations to more general boundary conditions, to transmission of transverse waves, and to separation of bulk and surface contributions. A generalization of the present calculation to take Fermi-liquid effect into account in the superfluid state is under consideration.

\begin{acknowledgments}
We thank Andrew Casey, William Halperin, Yoonseok Lee, Katsuhiko Nagai, Ryuji Nomura, Yuichi Okuda, Jeevak Parpia, James Sauls,  John Saunders, and Priya Sharma for useful discussions. We thank Esa Kivirinta for computing the results of Refs.\ \onlinecite{Flowers78,Richardson78}.
 This work was supported by the Academy of Finland and Tauno T\"onning foundation.
 \end{acknowledgments}

\end{document}